\documentclass[twocolumn,showpacs,amsmath,amssymb]{revtex4}
\usepackage[dvips]{graphics}
\def\imo{i}
\def\re#1{Re(#1)}
\def\im#1{Im(#1)}

\begin{document}
\title{Radiation processes in the vicinity of non-Schwarzschild and non-Kerr black holes}
\author{R. A. Konoplya}\email{konoplya_roma@yahoo.com}
\affiliation{DAMTP, Centre for Mathematical Sciences, University of Cambridge, Wilberforce Road, Cambridge CB3 0WA, UK}
\author{A. Zhidenko}\email{olexandr.zhydenko@ufabc.edu.br}
\affiliation{Centro de Matem\'atica, Computa\c{c}\~ao e Cogni\c{c}\~ao, Universidade Federal do ABC (UFABC), Rua Aboli\c{c}\~ao, CEP: 09210-180, Santo Andr\'e, SP, Brazil}
\begin{abstract}
  Usually alternative theories of gravity imply deviations from the well-known Kerr space-time, a model of an isolated black hole in General Relativity. In the dominant order, the deformed Kerr metric, free of closed time-like curves outside the event horizon, has been suggested recently by Johannsen and Psaltis. It has a single deformation parameter which is not constrained by the current observations, allowing, thereby, for a kind of unified and simple phenomenological description of black holes in various theories of gravity. Here we consider a number of classical and quantum phenomena of radiation in the vicinity of such deformed Schwarzschild-like and Kerr-like black holes: spiralling of particles into black holes, decay of fields propagating in the black hole's background, Hawking radiation. In particular, we calculate some quantitative characteristics of the above phenomena, such as the binding energy of particles, quasinormal modes, late-time tails of fields of various spin, intensity of Hawking radiation. The binding energy released when a particle goes over from a given stable orbit in the equatorial plane to the innermost stable one is calculated for such non-Kerr black holes. Due to inseparability of the wave equations in the general case, the perturbations and stability of scalar, Dirac, and electromagnetic fields are analyzed for vanishing rotation only. The dependence of the radiation phenomena on the deformation parameter is discussed.
\end{abstract}
\maketitle

\section{Introduction}

Rotating black holes in 3+1 General Relativity are described by the well-known Kerr solution, which is singled out by the uniqueness theorem: Kerr space-time is the only stationary, axisymmetric, asymptotically flat solution to the Einstein equations in vacuum, which possesses an event horizon and free of closed time-like curves outside of it \cite{Robinson:1975bv}. Thus, only mass $M$ and angular momentum $L$, besides the electric charge $Q$ which must be tiny for large astrophysical black holes, completely determine the properties of black holes in General Relativity. Therefore, Kerr space-time has been extensively studied since its discovery in the context of various potentially observable phenomena, such as radiation of gravitational waves, accretion of matter, gravitational lensing, and also in the scope of theoretical questions related to evolution of black holes, gravitational collapse, Hawking radiation and others \cite{BH}.

At the same time, Einstein's General Relativity, being apparently the simplest geometric theory of gravitation among others, raises a number of fundamental questions concerning the nature of singularities, dark energy and dark matter, quantization of gravitational interactions. Attempts to resolve these questions lead theorists to a number of alternative theories of gravity (e.g. the so-called modified theories of gravity, gravities with Chern-Simons', higher-curvature, and dilaton terms, brane-world scenarios, etc), which give the same observational consequences at the modern level of experiments, because Einstein's gravity, as well as a number of its alternatives, has been tested only in the regime of the post-Newtonian approximations, but non as a fully non-linear theory. A future experimental test of the strong field regime may be given by gravitational interferometers, which could detect gravitational
waves from neutron stars and black holes through observation of the quasinormal modes of these compact objects. Recently a number of works have been devoted to the study of potentially observable processes around various non-Kerr black holes \cite{Johannsen:2010xs,Hughes:2010xf,Apostolatos:2009vu,Bambi:2011yz,Pani:2011gy}. Therefore, a unified description of analogues of the Schwarzschild and Kerr solutions in various theories of gravity would be most useful for testing the alternatives. Such a model was suggested by Johannsen and Psaltis \cite{Joh_Psal} who considered deviations from the Schwarzschild and Kerr solutions and found a regular outside the event horizon space-time, which reduces to the Kerr one, when the deformation parameters vanish.
The Johannsen and Psaltis metric is not a vacuum solution of the Einstein equations, but is obtained in a kind of perturbative way in order to include various possible deviations from the Kerr solution in alternative theories of gravity (notice that an earlier attempt in constructing of the non-Kerr solutions was made by Glampedakis and Babak \cite{Glampedakis:2005cf}).
Current observations restrain values of the deformation parameters $\varepsilon_i$, so that in the dominant order only one deformation parameter $\varepsilon_3=\varepsilon$ remains which leads either to more prolate ($\varepsilon >0$) or more oblate ($\varepsilon <0$) shape of the horizon than the one of Kerr. Thus, the Johannsen-Psaltis space-time is described by three parameters: mass $M$, angular momentum per unit of mass $a$ and deformation $\varepsilon$.

Recently the Johannsen-Psaltis background has been studied in a number of papers \cite{Joh_Psal_study} with a special emphasis on gravitational lensing and Penrose processes of energy extraction around such black holes.
Here, our aim is to study various processes of radiation (both classical and quantum) around such a generic black hole's model.
A classical radiation of a field near a black hole consists of three stages, at each of which one of the following processes is dominating:
\begin{enumerate}
  \item initial outburst, which crucially depends on the initial perturbation,
  \item the intermediate damped oscillations represented by complex frequencies, termed quasinormal modes, and
  \item the asymptotic tails at late times.
\end{enumerate}
Therefore, here we study both processes which are independent on the initial conditions of the perturbation, the quasinormal modes and late-time tails, though, in the limit of vanishing rotation, due to inseparability of variables in field equations for the most general case. The Hawking radiation, for estimation of which one requires the classical reflection index of fields, is also considered for the non-Schwarzschild case.

Another important characteristic which we consider is the binding energy of a particle moving from a given equatorial orbit to the innermost stable one. The binding energy allows one to learn how much energy the matter (for example an accretion disk) will release before plunging into the black hole. Here we compute the binding energy for arbitrary rotation and deformation parameters $a$ and $\varepsilon$. We have found that the binding energy is increasing if the shape of the horizon is more prolate than the one of Kerr solution, that is, when $\varepsilon<0$. The Hawking radiation, on the contrary, is enhanced for larger values of $\varepsilon$. Real oscillation frequencies of quasinormal ringing are increasing when the deformation parameter $\varepsilon$ grows.

The paper is organized as follows: In Sec II we give some basic properties of the Johannsen-Psaltis black hole which will be explored in Sec III, devoted to deducing the wave equations and stability of scalar, electromagnetic, and Dirac fields in the spherically symmetric background. Sec. IV describes numerical methods used in the paper. In order to find quasinormal modes we used Frobenius expansion \cite{Leaver:1985ax,Nollert}, WKB method \cite{WKB}, and time-domain integration \cite{LTT}, the last one was also used for getting late-time tails. Sec V summarize the obtained results on quasinormal modes and late-time tails, while in Sec VI the intensity of Hawking radiation is computed for fields of various spin.

Note that throughout the paper we use geometrical units $G=c=1$ and fix the scale symmetry in all the numerical calculations such that the black hole mass $M=1/2$.

\section{The non-Schwarzschild and non-Kerr backgrounds}

In Boyer-Lindquist coordinates the Johannsen-Psaltis metric \cite{Joh_Psal} can be written as
\begin{widetext}
\begin{eqnarray}
ds^2 &=& [1+h(r,\theta)] \left(1-\frac{2Mr}{\Sigma}\right)dt^2 +\frac{ 4aMr\sin^2\theta }{ \Sigma }[1+h(r,\theta)]dtd\phi -  \frac{ \Sigma[1+h(r,\theta)] }{ \Delta + a^2\sin^2\theta h(r,\theta) }dr^2
\nonumber\\&&
- \Sigma d\theta^2
- \left[ \sin^2\theta \left( r^2 + a^2 + \frac{ 2a^2 Mr\sin^2\theta }{\Sigma} \right) + h(r,\theta) \frac{a^2(\Sigma + 2Mr)\sin^4\theta }{\Sigma} \right] d\phi^2,
\label{metric}
\end{eqnarray}
\end{widetext}
where  $\Sigma = r^2 + a^2 \cos^2\theta$, $M$ is the black hole mass, $a$ is the rotation parameter, and the deformation parameter $\varepsilon$ comes from the following general expansion

\begin{equation}
h(r,\theta)\equiv \sum_{k=0}^\infty \left( \varepsilon_{2k} + \varepsilon_{2k+1}\frac{Mr}{\Sigma} \right) \left( \frac{M^2}{\Sigma} \right)^{k}.
\label{h(r,theta)}
\end{equation}

The requirement of the asymptotic flatness of the metric implies that $\varepsilon_0=\varepsilon_1=0$. The current Lunar Laser Ranging experiment constrains the post-Newtonian parameters of the metric, so that $\varepsilon_2$ can be neglected, and, in the dominant order, one can take \cite{Joh_Psal}:
\begin{equation}
h(r,\theta) = \varepsilon \frac{M^3 r}{\Sigma^2}.
\label{hchoice}
\end{equation}
Various astrophysical constraints on $\varepsilon$ can be also found in \cite{Bambi:2011ew,Bambi:2012zg}.

In the next section we shall use the coefficients of the inverse metric in the equatorial plane ($\theta=\pi/2$):
\begin{widetext}
\begin{equation}
g^{\mu \nu} =
\left(
\begin{array}{llll}
 \frac{r^2 \left(r^6+a^2 (2 M+r) \left(\varepsilon  M^3+r^3\right)\right)}{\left(\varepsilon  M^3+r^3\right) \left((r-2 M) r^4+a^2 \left(\varepsilon
   M^3+r^3\right)\right)} & 0 & 0 & \frac{2 a M r^2}{(r-2 M) r^4+a^2 \left(\varepsilon  M^3+r^3\right)} \\
 0 & -\left(\frac{a^2}{r^2}+\frac{r^2 (r-2 M)}{\varepsilon  M^3+r^3}\right) & 0 & 0 \\
 0 & 0 & -\frac{1}{r^2} & 0 \\
 \frac{2 a M r^2}{(r-2 M) r^4+a^2 \left(\varepsilon  M^3+r^3\right)} & 0 & 0 & -\frac{r^2 (r-2 M)}{(r-2 M) r^4+a^2 \left(\varepsilon  M^3+r^3\right)}
\end{array}
\right).
\end{equation}
\end{widetext}
The event horizon is located at the corresponding root of the equation $g_{t\phi}^2 - g_{tt} g_{\phi\phi} = 0$, and the deformation parameter $\varepsilon$, once positive (negative) leads to a more prolate (oblate) object than the Kerr black hole.
As $g_{\theta\theta}>0$ and $g_{rr}>0$, $g_{\phi\phi}>0$, this space-time is free of closed timelike curves. For arbitrary $-8<\varepsilon\leq 0$ the space-time has the closed event horizon and represents therefore a black hole. On the contrary, $\varepsilon >0$ constrains the maximal value of the rotation parameter $a$ (see Fig.~2 in \cite{Joh_Psal}), which becomes smaller than the one of Kerr solution.

The case $\varepsilon\leq -8$ corresponds to a very large deformation, which is not observed in nature. Therefore, we shall not consider
radiation phenomena for  $\varepsilon\leq -8$. Yet, let us briefly comment on behavior of the  Johannsen-Psaltis metric in this regime.
The solution to equation $1+h(r,\theta)=0$ corresponds to the space-time singularity, since the curvature and the Kretschmann invariant approach infinity there as
\begin{eqnarray}
R&\propto&(1+h(r,\theta))^{-3}, \quad 1+h(r,\theta) \rightarrow 0,\\\nonumber
R_{\mu\nu\sigma\rho}R^{\mu\nu\sigma\rho}&\propto&(1+h(r,\theta))^{-6}, \quad 1+h(r,\theta) \rightarrow 0.
\end{eqnarray}
At the same time,  equation $1+h(r,\theta)=0$ at $\varepsilon\leq -8$ corresponds to an event
horizon which is situated outside the Kerr horizon. Thus, the metric is not regular at the outer event horizon for $\varepsilon \leq -8$,
on the contrary to the claim of \cite{Joh_Psal} (see, for example, fig. 2 in \cite{Joh_Psal}).

When $\varepsilon <-8$ the above singular outer event horizon can be reached by an in-falling photons in finite time according to the remote observer's clock, as
$$\theta,\phi=const,\quad ds=0\quad\Longrightarrow\quad dt=\left(1-\frac{2Mr}{\Sigma}\right)^{-1}dr,$$
and  the integral remains finite, when integrated from the outer event horizon:
$$T=\int dt=\int\left(1-\frac{2Mr}{\Sigma}\right)^{-1}dr<\infty.$$

\section{Binding energy}

\begin{figure*}
\resizebox{\linewidth}{!}{\includegraphics*{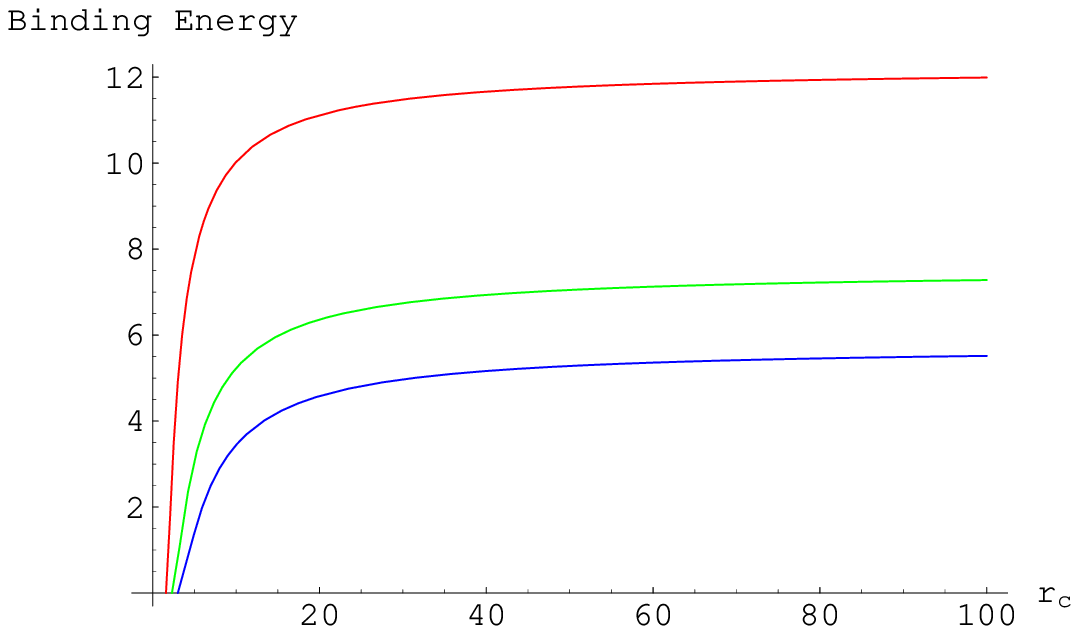}\includegraphics*{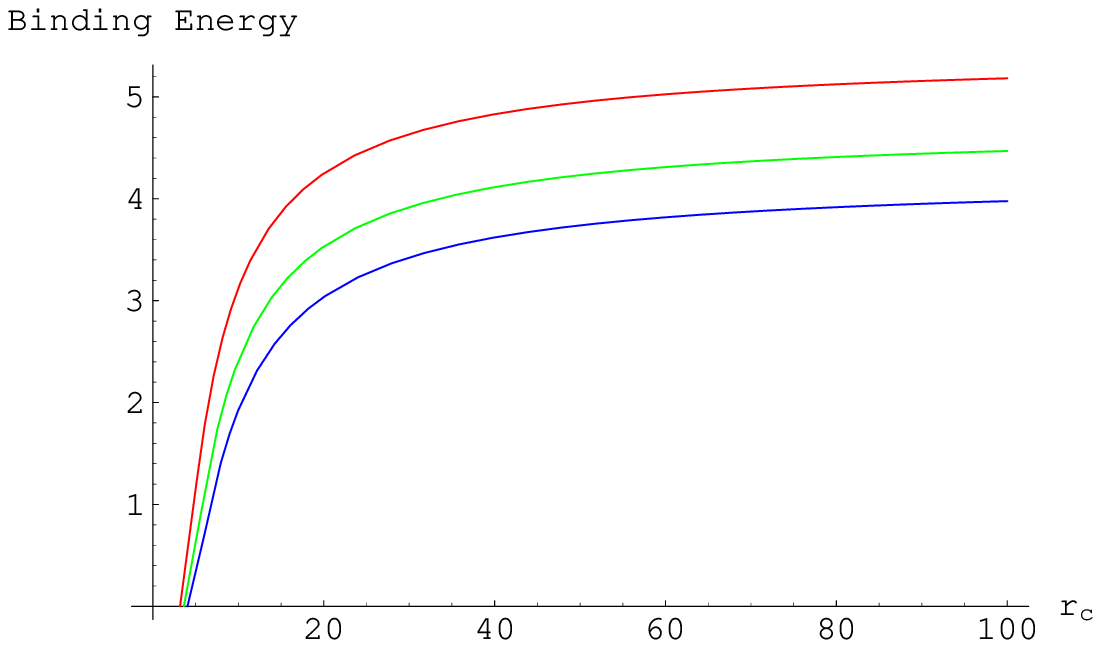}}
\caption{1) Left panel: Binding energy as a function of $r_c$ for $\varepsilon = 5$ (top), $0.01$, and $-5$ (bottom), $a=0.2$ for particles co-rotating with a black hole at a circular orbit $r_{c}$. 2) Right panel: Binding energy for $\varepsilon = 5$ (top), $0.01$, and $-5$ (bottom), $a=0.2$ for counter-rotating particles at a circular orbit $r_{c}$.}
\label{BE-plots}
\end{figure*}

In order to learn how much energy could be extracted from a particle slowly spiralling toward the black hole's horizon, it is necessary to learn what the energy of the particle at the innermost stable circular orbit is.  For this purpose it is useful to introduce a function termed the \emph{binding energy}, which is the amount of energy released by the particle going over from a given stable circular orbit located at $r_c$ to the innermost stable orbit $r_{is}$. Thus, the binding energy in percent is
\begin{equation}
\text{Binding Energy}  = 100\times \frac{E(r_c)- E(r_{is})}{E(r_c)}.
\label{BE}
\end{equation}
Frequently, instead of some given circular orbit $r_c$ a particle coming from infinity is considered, so that $r_c = \infty$ is taken in the above definition of the binding energy. The binding energy for the Schwarzschild black hole is about $5.7$ percent and for the extremal Kerr black hole can be as large as $42$ percent. This is much larger than, for instance, the binding energy at the nuclear fusion, which is about $0.7$ percent of the rest mass energy. Recently the binding energy for charged particles around a black hole deformed by the tidal force and magnetic field has been found in \cite{Konoplya:2012vh,Konoplya:2006qr}.

For our case of a stationary, axisymmetric metric, one has three integrals of the motion,
\begin{equation}
p^\alpha = \mu \frac{dx^\alpha}{d\tau},
\end{equation}
which are particle's rest mass $\mu$, energy $E=p_t$, and angular momentum $L=-p_\phi$.
Then, the equation
\begin{equation}
p_\alpha p^\alpha = \mu^2
\label{equation1}
\end{equation}
leads in the equatorial plane to the following relation:
\begin{equation}
\left(\frac{dr}{d\tau}\right)^2 = -\frac{1}{g_{rr}} \left( g^{tt}E^2 - 2g^{t\phi}E L + g^{\phi\phi}L^2 - \mu^2 \right).
\end{equation}
The energy $E$ and momentum $L$ of a particle at a circular orbit $r$ can be determined numerically from the requirements:
\begin{equation}
\frac{d r}{d \tau} =0, \quad \frac{d^2 r}{d \tau^2} =0.
\end{equation}
Then, the numerically obtained $E(r, a, \varepsilon)/\mu$, $L(r, a, \varepsilon)/\mu$ allow one to use formula (\ref{BE}).

\begin{table}
\caption{Binding energy $B_e$ for various values of the deformation parameter $\varepsilon$  and rotation $a$. Left columns are for particles co-rotating with a black hole, right columns are for counter rotating particles. The star means that the extremal values $a =0.5$ for co-rotating particles was reached in computation only asymptotically, as the innermost stable orbit approaches the horizon in this case.}
\label{Table-BE}
\begin{tabular}{|r|l|c|c|c|c|c|c|}
  \cline{3-8}
  \multicolumn{2}{c|}{}&\multicolumn{3}{|c|}{co-rotating}&\multicolumn{3}{|c|}{counter rotating}\\
  \hline
  $\varepsilon$ & $a$ & $r_{is}$ & $E(r_{is})$ & $B_e$ 
  & $r_{is}$ & $E(r_{is})$ & $B_e$ \\
  \hline
  -5 & 0.1 & 3.26251 & 0.947203 & 5.27972 
  & 3.78356 & 0.954798 & 4.52022 \\
  \hline
  -5 & 0.2 & 3.01354 & 0.942488 & 5.75122 
  & 4.04775 & 0.957840 & 4.21596 \\
  \hline
  -5 & 0.3 & 2.78166 & 0.937117 & 6.28830 
  & 4.31171 & 0.960488 & 3.95116 \\
   \hline
  -5 & 0.4 & 2.57733 & 0.931213 & 6.87875 
  & 4.57452 & 0.962809 & 3.71907 \\
   \hline
  -5 & 0.5$^*$ & 2.41140 & 0.925130 & 7.48702 
  & 4.83569 & 0.964858 & 3.51420 \\
   \hline
  -1 & 0.1 & 2.79356 & 0.938409 & 6.15909 
  & 3.41641 & 0.949953 & 5.00469 \\
   \hline
  -1 & 0.2 & 2.46574 & 0.929858 & 7.01424 
  & 3.71439 & 0.954065 & 4.59345 \\
  \hline
  -1 & 0.3 & 2.12623 & 0.917971 & 8.20290 
  & 4.00518 & 0.957474 & 4.25260 \\
   \hline
  -1 & 0.4 & 1.78144 & 0.900507 & 9.94928 
  & 4.28970 & 0.960353 & 3.96472 \\
  \hline
  -1 & 0.5$^*$ & 1.46393 & 0.874561 & 12.5439 
  & 4.56904 & 0.962822 & 3.71779 \\
  \hline
  1 & 0.1 & 2.53274 & 0.931827 & 6.81727 
  & 3.22108 &  0.946834  & 5.31664 \\
  \hline
  1 & 0.2 & 2.14187 & 0.918816 & 8.11845 
  & 3.53857 &  0.951732 &  4.82680 \\
  \hline
  1 & 0.3 & 1.67902 & 0.895150 & 10.4850 
  & 3.84436 &  0.955666  & 4.43339 \\
  \hline
  5 & 0.1 & 2.04540 & 0.911207 & 8.87926 
  &  2.82637 &  0.938547 &  6.14526 \\
  \hline
  5 & 0.2 & 1.59086 & 0.877896 & 12.2104 
  & 3.17842 &  0.945817 &  5.41832 \\
  \hline
  5 & 0.25 & 1.30398 & 0.841900 & 15.8100 
  & 3.34747 &  0.948706 &  5.12938 \\
  \hline
\end{tabular}
\end{table}

On Fig.~\ref{BE-plots} one can see that the more prolate (or the less oblate) shape of the horizon is, the larger the binding energy is. This is true for any values of the rotation parameter $a$, as it is shown on table~\ref{Table-BE}. An increase in the binding energy is slightly larger for highly rotating black holes than for slowly rotating ones and can be about 100 percent for relatively small deformations. Indeed, $\varepsilon = 5$ means deformation of order $\varepsilon M^3/r^3 \sim 5/8$ and corresponds to approximately $2.5$  times increase (see table~\ref{Table-BE}) in the binding energy.

We know that the binding energy for the extremal Kerr black hole equals $3.8$ percent for counter-rotating particles and $42$ percent for co-rotating ones. In concordance with this we obtain in the limit of small $\varepsilon$ for the non-Kerr black hole: for $a=0.4999999$, $\varepsilon = -10^{-6}$, $E_b = 41.46$ percent for co-rotating particle and $E_b = 3.78$ percent for counter rotating one. Thus, in the limit of vanishing deformation parameter the binding energy approaches its Kerr values.

\section{Deduction of the master wave equations for scalar, electromagnetic, and Dirac fields}

In the most general case we were unable to decouple variables in the field equation, and had to be limited by the non-rotating case.
Although scalar and electromagnetic wave equations can be easily derived for this case, the Dirac field requires some
algebraic calculations.
Therefore, we deduce here the wave equations for spin $0$, $\pm 1/2$ and $\pm 1$ fields in the general spherically symmetric background using the tetrad formalism.

\subsection{Spherically symmetric backgrounds and the tetrad formalism}

A spherically symmetric background can be described by the following line element
\begin{equation}\label{ss-ll}
ds^2 = g_{\mu\nu}dx^\mu dx^\nu = Adt^2-\frac{B^2}{A}dr^2-r^2(d\theta^2+\sin^2\theta d\varphi^2),
\end{equation}
where the metric coefficients $A$ and $B$ depend in general case on both $t$ and $r$ coordinates.

The scalar field satisfies the Klein-Gordon equation,
\begin{equation}\label{KG}
\frac{1}{\sqrt{-g}}\frac{\partial}{\partial x^\mu}\left(g^{\mu\nu}\sqrt{-g}\frac{\partial\Psi}{\partial x^\mu}\right)=0.
\end{equation}

The massless Dirac and Maxwell equations are
\begin{eqnarray}
\label{Dirac}&&
\begin{array}{rcl}\sqrt{2}\nabla_{BB'}P^B&=&0,\\
\sqrt{2}\nabla_{BB'}Q^B&=&0,
\end{array}\\
\label{Maxwell}&&F^{\mu\nu}_{~~;\mu}=0\qquad (F_{\mu\nu}=\partial_\mu A_\nu-\partial_\nu A_\mu),
\end{eqnarray}
where ${}_{;\mu}$ and $\nabla_{BB'}$ denote covariant differentiation.

Following \cite{Newman:1961qr} we rewrite (\ref{Dirac}) and (\ref{Maxwell}) in a tetrad system of null-vectors:
\begin{equation}
\begin{array}{rclrcl}
l_\mu&=&\displaystyle\frac{1}{2}\{A,B,0,0\},&\quad m_\mu&=&\displaystyle\frac{r}{\sqrt{2}}\{0,0,1,\imo\sin\theta\},\\
n_\mu&=&\displaystyle\{1,-\frac{B}{A},0,0\},&\quad \overline{m}_\mu&=&\displaystyle\frac{r}{\sqrt{2}}\{0,0,1,-\imo\sin\theta\},
\end{array}
\end{equation}
that satisfy the following relations
\begin{eqnarray}
\nonumber&l_\mu l^\mu=n_\mu n^\mu=m_\mu m^\mu=\overline{m}_\mu \overline{m}^\mu=0,&\\
\nonumber&l_\mu m^\mu=l_\mu \overline{m}^\mu=n_\mu m^\mu=n_\mu \overline{m}^\mu=0,&\\
\nonumber&l_\mu n^\mu=1,\qquad m_\mu \overline{m}^\mu=-1,&\\
\nonumber&l_\mu n_\nu+n_\mu l_\nu-m_\mu \overline{m}_\nu-\overline{m}_\mu m_\nu=g_{\mu\nu}.&
\end{eqnarray}
The spin coefficients are
\begin{eqnarray}
\nonumber&\displaystyle\kappa=l_{\mu;\nu}m^\mu l^\nu = 0, \qquad \pi=-n_{\mu;\nu}\overline{m}^\mu l^\nu = 0, &\\
\nonumber&\displaystyle\epsilon=\frac{1}{2}(l_{\mu;\nu}n^\mu l^\nu-m_{\mu;\nu}\overline{m}^\mu l^\nu) = \frac{\dot{B}-A^\prime}{4B},&\\
\nonumber&\displaystyle\rho=l_{\mu;\nu}m^\mu \overline{m}^\nu=\frac{A}{2rB},\qquad\lambda=-n_{\mu;\nu}\overline{m}^\mu \overline{m}^\nu=0\\
\label{spin-coefficients}&\displaystyle\alpha=\frac{1}{2}(l_{\mu;\nu}n^\mu \overline{m}^\nu-m_{\mu;\nu}\overline{m}^\mu \overline{m}^\nu)=\frac{\cot\theta}{2\sqrt{2}r},&\\
\nonumber&\displaystyle\displaystyle\sigma=l_{\mu;\nu}m^\mu m^\nu=0,\qquad\mu=-n_{\mu;\nu}\overline{m}^\mu m^\nu=\frac{1}{rB},&\\
\nonumber&\displaystyle\beta=\frac{1}{2}(l_{\mu;\nu}n^\mu m^\nu-m_{\mu;\nu}\overline{m}^\mu m^\nu)=-\frac{\cot\theta}{2\sqrt{2}r},&\\
\nonumber&\displaystyle\nu=-n_{\mu;\nu}\overline{m}^\mu n^\nu=0,\qquad\tau=l_{\mu;\nu}m^\mu n^\nu=0&\\
\nonumber&\displaystyle\gamma=\frac{1}{2}(l_{\mu;\nu}n^\mu n^\nu-m_{\mu;\nu}\overline{m}^\mu n^\nu)=\frac{\dot{A}}{2A^2}-\frac{\dot{B}}{2AB},&
\end{eqnarray}
where dot and prime denote derivatives with respect to $t$ and $r$.

The intrinsic derivatives read
\begin{equation}\label{intrinsic-derivatives}
\begin{array}{l}
\displaystyle D={}_{;\mu}l^\mu=\frac{1}{2}\frac{\partial}{\partial t}-\frac{A}{2B}\frac{\partial}{\partial r},\\
\displaystyle\Delta={}_{;\mu}n^\mu=\frac{1}{A}\frac{\partial}{\partial t}-\frac{1}{B}\frac{\partial}{\partial r},\\
\displaystyle\delta={}_{;\mu}m^\mu=-\frac{1}{\sqrt{2}r}\frac{\partial}{\partial \theta}-\frac{\imo}{\sqrt{2} r\sin\theta}\frac{\partial}{\partial \varphi},\\
\displaystyle\overline{\delta}={}_{;\mu}\overline{m}^\mu=-\frac{1}{\sqrt{2}r}\frac{\partial}{\partial \theta}+\frac{\imo}{\sqrt{2} r\sin\theta}\frac{\partial}{\partial \varphi}.
\end{array}
\end{equation}

In terms of these variables the massless Dirac equations can be reduced to \cite{Page:1976}:
\begin{equation}\label{Dirac-NP}
\begin{array}{rcl}
(D+\epsilon-\rho)P^0+(\overline{\delta}+\pi-\alpha)P^1&=&0,\\
(\Delta+\mu-\gamma)P^1+(\delta+\beta-\tau)P^0&=&0,\\
(D+\overline{\epsilon}-\overline{\rho})\overline{Q}^0+(\delta+\overline{\pi}-\overline{\alpha})\overline{Q}^1&=&0,\\
(\Delta+\overline{\mu}-\overline{\gamma})\overline{Q}^1+(\overline{\delta}+\overline{\beta}-\overline{\tau})\overline{Q}^0&=&0.
\end{array}
\end{equation}
The Maxwell equations take the form \cite{Newman:1961qr}:
\begin{equation}\label{Maxwell-NP}
\begin{array}{rcl}
D\Phi_1-\overline{\delta}\Phi_0&=&(\pi-2\alpha)\Phi_0+2\rho\Phi_1-\kappa\Phi_2,\\
D\Phi_2-\overline{\delta}\Phi_1&=&-\lambda\Phi_0+2\pi\Phi_1+(\rho-2\epsilon)\Phi_2,\\
\delta\Phi_1-\Delta\Phi_0&=&(\mu-2\gamma)\Phi_0+2\tau\Phi_1-\sigma\Phi_2,\\
\delta\Phi_2-\Delta\Phi_1&=&-\nu\Phi_0+2\mu\Phi_1+(\tau-2\beta)\Phi_2,\\
\end{array}
\end{equation}
where $\Phi_0=F_{\mu\nu}l^\mu m^\nu$, $\Phi_1=\frac{1}{2}F_{\mu\nu}(l^\mu n^\nu+\overline{m}^\mu m^\nu)$, and $\Phi_2=F_{\mu\nu}\overline{m}^\mu n^\nu$.

Substituting the following ansatz
\begin{equation}\label{ansatz}
\begin{array}{rcl}
\Psi&=&R_0(t,r){~}_{0}Y_{\ell m}(\theta,\varphi)r^{-1},\\
P^0&=&R_{+1/2}(t,r){~}_{+1/2}Y_{\ell m}(\theta,\varphi)r^{-3/2}A(t,r)^{-1/4},\\
P^1&=&R_{-1/2}(t,r){~}_{-1/2}Y_{\ell m}(\theta,\varphi)r^{-3/2}A(t,r)^{1/4},\\
\overline{Q}^0&=&R_{+1/2}(t,r){~}_{-1/2}Y_{\ell m}(\theta,\varphi)r^{-3/2}A(t,r)^{-1/4},\\
\overline{Q}^1&=&R_{-1/2}(t,r){~}_{+1/2}Y_{\ell m}(\theta,\varphi)r^{-3/2}A(t,r)^{1/4},\\
\Phi_0&=&R_{-1}(t,r){~}_{-1}Y_{\ell m}(\theta,\varphi)r^{-2}A(t,r)^{1/2},\\
\Phi_2&=&R_{+1}(t,r){~}_{+1}Y_{\ell m}(\theta,\varphi)r^{-2}A(t,r)^{-1/2},
\end{array}
\end{equation}
into (\ref{KG},\ref{Dirac-NP},\ref{Maxwell-NP}) (here, ${}_sY_{\ell m}(\theta,\phi)$ are the spin-weighted spherical harmonics with $\ell=|s|,|s|+1,|s|+2,\ldots$ and $m=-\ell,-\ell+1,\ldots\ell-1,\ell$), we
obtain the following second order equation for $R_s(t,r)$:
\begin{widetext}
\begin{eqnarray}
\nonumber&&\frac{A}{B}\frac{\partial}{\partial r}\left(\frac{A}{B}\frac{\partial R_s}{\partial r}\right)+s\frac{\dot{A}B-2A\dot{B}}{B^2}\frac{\partial R_s}{\partial r}
-\frac{A}{B}\frac{\partial}{\partial t}\left(\frac{B}{A}\frac{\partial R_s}{\partial t}\right)
-s\frac{2A-rA^\prime}{rB}\frac{\partial R_s}{\partial t}-\left(\frac{A\overline{\lambda}}{r^2}+\frac{AA^\prime}{rB^2}-\frac{A^2B^\prime}{rB^3}\right)R_s
\\\label{master-equation}&&+|s|\left(\frac{A^2}{r^2B^2}+\frac{2AA^\prime}{rB^2}-\frac{2A^2B^\prime}{rB^3}-\frac{AA^\prime B^\prime}{2B^3}+\frac{AA^{\prime\prime}}{2B^2}-\frac{\dot{A}^2}{A^2}+\frac{3\dot{A}\dot{B}}{2AB}+\frac{\ddot{A}}{2A}-\frac{\ddot{B}}{B}\right)R_s
\\\nonumber&&-s^2\left(\frac{A^2}{r^2B^2}+\frac{AA^\prime}{rB^2}+\frac{{A^\prime}^2}{4B^2}-\frac{2A^2B^\prime}{rB^3}-\frac{\dot{A}^2}{4A^2}+\frac{\dot{A}\dot{B}}{AB}-\frac{\dot{B}^2}{B^2}\right)R_s
+s\left(\frac{A\dot{A}^\prime-\dot{A}A^\prime}{AB}+\frac{A(\dot{B}B^\prime-B\dot{B}^\prime)}{B^3}\right)R_s=0,
\end{eqnarray}
\end{widetext}
where $\overline{\lambda}=(\ell+|s|)(\ell-|s|+1)$.

Although we have derived equation (\ref{master-equation}) for $s=0,\pm1/2,\pm1$, we will use (\ref{master-equation})  only for scalar and Dirac fields, while for electromagnetic perturbations we shall use the following simpler equation,
\begin{equation}
\frac{A}{B}\frac{\partial}{\partial r}\left(\frac{A}{B}\frac{\partial {\tilde R}_{\pm1}}{\partial r}\right)
-\frac{A}{B}\frac{\partial}{\partial t}\left(\frac{B}{A}\frac{\partial {\tilde R}_{\pm1}}{\partial t}\right)
-\frac{A\overline{\lambda}{\tilde R}_{\pm1}}{r^2}=0,
\end{equation}
which can be derived from either the first and third or the second and forth equations of (\ref{Maxwell-NP}) by the substitution
\begin{equation}
\Phi_1={\tilde R}_{\pm1}(t,r){~}_{\pm1}Y_{\ell m}(\theta,\varphi)r^{-2}.
\end{equation}

\subsection{Linear perturbation equations and stability of fields in the non-Schwarzschild black hole background}

\begin{figure*}
\resizebox{\linewidth}{!}{\includegraphics*{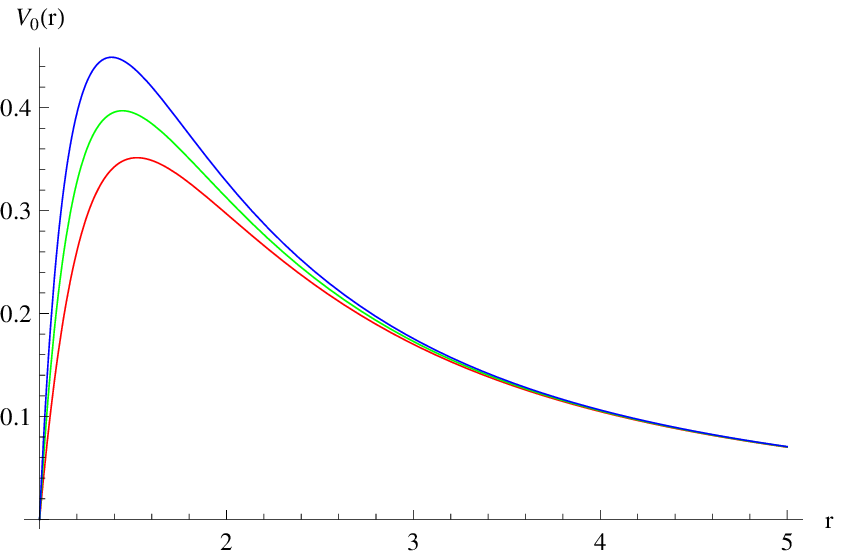}\includegraphics*{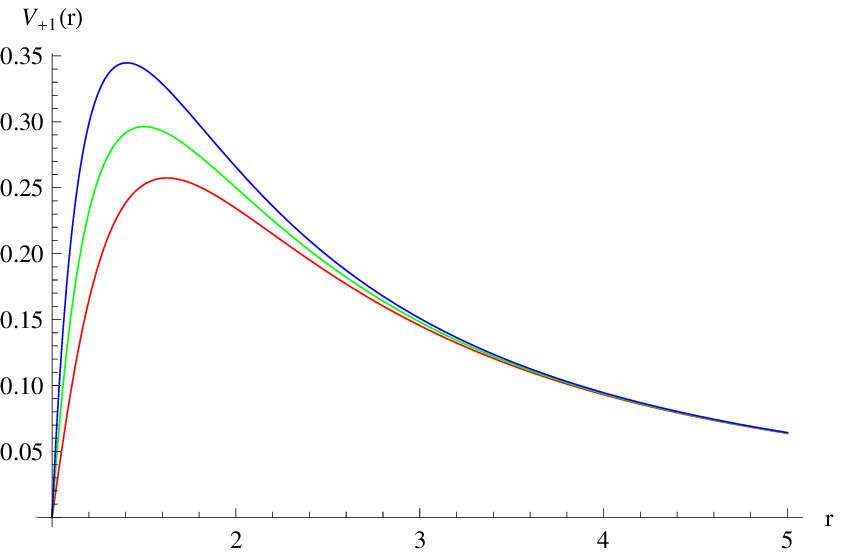}}
\caption{The effective potentials ($\ell=1$) for the scalar (left panel) and Maxwell (right panel) fields for $\varepsilon=-4$ (red, lower), $\varepsilon=0$ (green) and $\varepsilon=4$ (blue, upper).}
\label{potentialsQNMs}
\end{figure*}

The Schwarzschild-like metric takes the form
\begin{equation}\label{Schwarzschild-like}
ds^2=(1+h(r))\left(f(r)dt^2-\frac{dr^2}{f(r)}\right)-r^2(d\theta^2+\sin^2\theta d\varphi^2),
\end{equation}
where $\displaystyle f(r)=1-\frac{2M}{r}$, $\displaystyle h(r)=\varepsilon\frac{M^3}{r^3}$.

Making use of the relations $A=(1+h(r))f(r)$, $B=1+h(r)$, and $R_s(t,r)=e^{-\imo\omega t}\Psi_s(r)$ in (\ref{master-equation}),
we find the wave-like equation
\begin{equation}\label{wavelike}
\frac{d^2\Psi_s}{dr_\star^2}+(\omega^2-V_s(r_\star))\Psi_s=0,
\end{equation}
where
$$dr_\star=\frac{dr}{f(r)}$$
is the tortoise coordinate. The effective potential $V_s$ for scalar, Dirac, and Maxwell fields has the following form
\begin{eqnarray}\label{effpot}
&\displaystyle V_0=k\overline{\lambda}+\frac{f(r)f'(r)}{r},\qquad V_{+1}=k\overline{\lambda},&\\\nonumber
&\displaystyle V_{+\frac{1}{2}}=k\overline{\lambda}+\frac{\imo\omega}{2k}\frac{dk}{dr_\star}-\frac{1}{4}\frac{d}{dr_\star}\left(\frac{1}{k}\frac{dk}{dr_\star}\right)+\frac{1}{16k^2}\left(\frac{dk}{dr_\star}\right)^2,\!\!&
\end{eqnarray}
where $\displaystyle k=\frac{1+h(r)}{r^2}f(r)$. Note that for $h(r)=0$ the potential $V_{+\frac{1}{2}}$ coincides with (2.17) of \cite{Jing:2005dt}. We will consider only $s\geq0$ because of the symmetry between opposite polarizations in a spherically symmetric background.

The effective potential for scalar and Maxwell fields has the form of the potential barrier that vanishes at the event horizon and infinity and is positive definite everywhere outside the event horizon (see fig.~\ref{potentialsQNMs}). Therefore, the solutions to the wave-like equations are eigenvectors of a positive self-adjoint operator in the Hilbert space of square integrable functions, implying that there are no growing modes in the quasinormal spectrum.

The effective potential for Dirac fields is complex due to its dependence on $\imo\omega$. However, similarly to the Schwarzschild black hole, one can introduce an equivalent wave-like equation with an effective potential that contains a square root. Following \cite{Khanal:1983vb}, we define a new function $Z$, such that
\begin{equation}
k^{1/4}\Psi_{+\frac{1}{2}}=\left(\sqrt{\overline{\lambda} k}-\imo\omega\right)Z+\frac{dZ}{dr_\star}.
\end{equation}
Substituting $\Psi_{+\frac{1}{2}}$ into the wave-like equation (\ref{wavelike}) we find that the new function $Z$ satisfies the wave-like equation with a real effective potential
\begin{equation}\label{25}
\frac{d^2Z}{dr_\star^2}+\left(\omega^2-\overline{\lambda} k+\frac{d\sqrt{\overline{\lambda} k}}{dr_\star}\right)Z=0.
\end{equation}
The corresponding differential operator is also positive self-adjoint one what can be shown with the help of $S$-deformation \cite{Kodama:2003kk} by taking $S=\sqrt{\overline{\lambda}k}$. Let us note, that we shall not use Eq.~(\ref{25}) for finding quasinormal modes, because the Frobenius method
cannot be easily applied to potentials with square roots \cite{Jing:2005dt,QNMs-Dirac-Carlu}.

Thus, the stability of massless fields in the nonrotating Johannsen-Psaltis background is straightforward \cite{Konoplya:2011qq} and basically identical to the Schwarzschild case.

The potential barrier (fig.~\ref{potentialsQNMs}) is growing when the deformation $\varepsilon$ is increasing. As we shall see later this feature of the effective potential will stipulate the character of dependence of the quasinormal modes and scattering amplitudes on $\varepsilon$.

\section{Numerical methods}

In this section we briefly discuss the numerical methods used for calculations of the quasinormal modes, the transmission/reflection coefficients, and late-time tails.

\subsection{Quasinormal modes and the Frobenius method}

In order to calculate quasinormal modes we impose the \emph{quasinormal mode boundary conditions} for the wave equation (\ref{wavelike}),
that is, we require that there are only purely ingoing waves at the black hole's horizon,
$$\Psi(r_*\rightarrow-\infty)\propto \exp(-\imo\omega r_*),$$
and only purely outgoing waves at spatial infinity, i.e.
$$\Psi(r_*\rightarrow\infty)\propto \exp(+\imo\omega r_*).$$
Thus, no waves are coming from the horizon or infinity, which implies that $\omega$ are proper oscillation modes in the black hole response to an ``instantaneous'' perturbation. In other words, when the perturbation decays, the source of the initial perturbation is not acting anymore.

Equation (\ref{wavelike}) has an irregular singularity at spatial infinity and a regular singular point at the horizon $r=2M$. The appropriate Frobenius series has the form
$$\Psi(r)=\left(\frac{r-2M}{r-R}\right)^{-\imo b}e^{\imo\omega r}(r-R)^{\imo c}\sum_{n=0}^{\infty}a_n\left(\frac{r-2M}{r-R}\right)^n,$$
where R is the closest to the horizon singular point. We find that $b=2M\omega$, $c=2M\omega$ for the scalar and Maxwell fields, while for the Dirac field $b=2M\omega-\imo/4$ and $c=2M\omega+\imo/2$, because of the $\omega$-dependent effective potential.

For the scalar and Maxwell fields the only other singularity is $r=0$ and, therefore, $R=0$. For the Dirac field there are additional singular points which are solutions to equation $1+h(r)=0$, so that $R=0$ for $\varepsilon\geq0$, and $R=M\sqrt[3]{-\varepsilon}$ for $\varepsilon<0$.

The coefficients $a_n$ satisfy a recurrence relation which can always be reduced to the three-term one through the Gaussian eliminations and, finally, one can find the equation with an infinite continued fraction (see e.g. \cite{Konoplya:2011qq} for details). This equation has an infinite number of roots $\omega_n$ corresponding to the quasinormal frequencies \cite{Leaver:1985ax}.  The infinite continued fraction converges very slowly when the imaginary part of $\omega$ is large. In this case, in order to improve the convergence, one can use the Nollert procedure \cite{Nollert}.

\subsection{Reflection coefficients}

In order to calculate the Hawking emission rates of particles, one needs to solve the problem of classical scattering and find the gray-body factors of the corresponding fields. Such a problem implies classical \emph{scattering boundary conditions} for the wave equations obtained in Sec III. Thus, at the event horizon the boundary condition corresponds to a purely ingoing wave, while at spatial infinity ($r\rightarrow\infty$) one has a sum of the ingoing and outgoing waves,
$$\Psi(r)\simeq Z_{in} \exp(-\imo\omega r_\star)+Z_{out} \exp(\imo\omega r_\star),$$
where $Z_{in}$ and $Z_{out}$ are integration constants. Thus, we would like to know which portion of particles will be able to pass through the barrier of the effective potential.

Introducing the new function
$$P(r)=\Psi(r)\left(\frac{r-2M}{r-R}\right)^{\imo b},$$
and choosing the integration constant as $P(2M)=1,$ we expand Eq. (\ref{wavelike}) near the event horizon and find $P'(2M)$, what completely fixes the initial conditions for the numerical integration. Then, we integrate Eq. (\ref{wavelike}) numerically from the event horizon $2M$ to some distant point $r_f\gg M$ and find a fit for the numerical solution far from the black hole in the following form:
\begin{equation}\label{fit}
P(r)=Z_{in} P_{in}(r)+Z_{out} P_{out}(r),
\end{equation}
where the asymptotic expansions for the corresponding functions are found by expanding (\ref{wavelike}) at large $r$ as
\begin{eqnarray}
P_{in}(r)&=&e^{-\imo\omega r}r^{-\imo c}\left(1+P_{in}^{(1)}r^{-1} + P_{in}^{(2)}r^{-2}+\ldots\right),\nonumber\\
P_{out}(r)&=&e^{\imo\omega r}r^{\imo c}\left(1+P_{out}^{(1)}r^{-1} + P_{out}^{(2)}r^{-2}+\ldots\right).\nonumber
\end{eqnarray}
The fitting procedure allows us to find $Z_{in}$ and $Z_{out}$. In order to check the accuracy of the found coefficients, one should increase the internal precision of the numerical integration, the value of $r_f$, and the number of terms in the series expansion for $P_{in}(r)$ and $P_{out}(r)$, making sure that the values of $Z_{in}$ and $Z_{out}$ do not change within desired precision. Thus, the above approach can be used for finding reflection coefficients of a wide class of compact objects  with great accuracy \cite{Kokkotas:2010zd}.

\subsection{WKB method}

In order to obtain an analytical expression for the quasinormal frequencies in the regime of large multipole numbers $\ell$ and also
for an additional check of the data obtained by the convergent Frobenius method, we shall use the WKB formula of the 6th order beyond the eikonal approximation \cite{WKB,WKBorder}. The formula has the following form:
\begin{equation}\label{WKBformula}
	\frac{\imo(\omega^2- V_{0})}{\sqrt{-2 V_{0}''}} - \sum_{i=2}^{i=6}
		\Lambda_{i} = n+\frac{1}{2},\qquad n=0,1,2\ldots,
\end{equation}
 where the correction terms $\Lambda_{i}$ were obtained in \cite{WKB,WKBorder} and depend on higher derivatives of $V$ at its maximum with respect to the tortoise coordinate $r_\star$, ($n$ labels the overtones). The WKB approach was developed by Schutz and Will \cite{WKB} and extended up to the 3rd   and 6th orders in \cite{WKBorder}. It can be effectively used not only for finding low-lying quasinormal modes, that is, oscillations with longer lifetime (see for instance \cite{WKBuse} and references therein), but also for calculations of the transmission/reflection coefficients in various problems \cite{Konoplya:2009hv,Konoplya:2010kv}.

\subsection{Time-domain integration}

In addition, we shall study late-time tails through the numerical characteristic integration method that uses the light-cone variables
$u = t - r^{*}$, $v = t + r^{*}$. In the characteristic initial value problem, initial data are specified on the two null surfaces $u = u_0$ and $v = v_0$. The discretization scheme was suggested in \cite{LTT} and used in a number of subsequent papers (see, for instance \cite{LTT-alot}), showing an excellent concordance with accurate Frobenius data at the stage of quasinormal ringing.

\section{Quasinormal modes and late-time tails}

\begin{figure*}
\resizebox{\linewidth}{!}{\includegraphics*{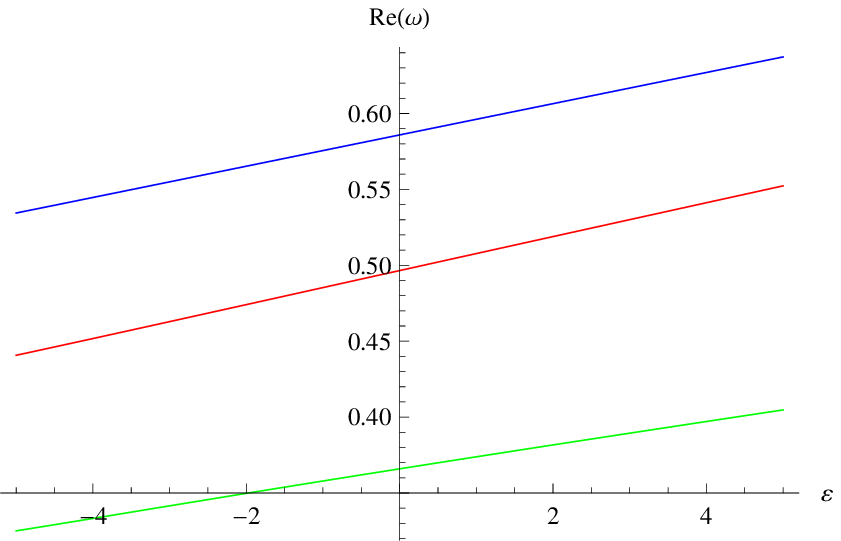}\includegraphics*{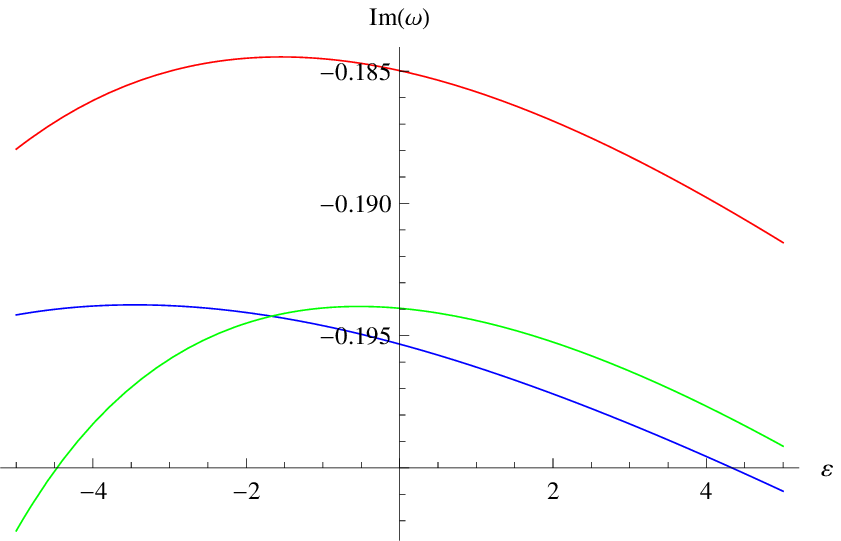}}
\caption{Real and imaginary parts of the fundamental quasinormal mode as a function of $\varepsilon$ for the scalar field $\ell=1$ (blue line, left panel top, right panel bottom), Dirac field $\ell=1/2$ (green line, left panel bottom), and Maxwell field $\ell=1$ (red line, right panel top) fields.}
\label{QNM}
\end{figure*}

Quasinormal modes can be written as
\begin{equation}
\omega_{n} = \re{\omega_{n}} + \imo\im{\omega_{n}},
\end{equation}
where $n$ is the overtone number. The real part of the frequency is the real oscillation frequency, while the imaginary part is proportional to the damping rate of the mode. In our designations, once $\im{\omega_{n}}$ is negative, the mode is damped, while positive imaginary part corresponds to growing (unstable) modes. As at late time the dominant in a signal mode corresponds to $n=0$, we shall consider here only this lowest mode, termed
the fundamental modes.

On Fig.~\ref{QNM} the fundamental quasinormal modes (the modes with the smallest decay rate) computed by the Frobenius method are shown for all the fields under consideration. A striking feature of the fundamental modes are almost linear dependence on $\varepsilon$ of their real oscillation frequencies. Possibly, this is connected with a relative smallness of the deformation $h$ for $\varepsilon$ ranging from $-5$ to $5$.  In the regime of strong deformations, that is certainly not justified physically, the dependence on $\varepsilon$ might be not linear anymore.
The fundamental modes can be approximated by the following analytic expressions for the scalar ($\ell=1$), Dirac ($\ell=1/2$), and Maxwell ($\ell=1$) fields respectively:
\begin{widetext}
\begin{equation}\label{eexp}
  \begin{array}{rcll}
    \omega & \approx & 0.58587(1+0.01756\varepsilon+0.00000\varepsilon^2)-0.19532\imo(1+0.00363\varepsilon+0.00045\varepsilon^2) & \qquad\mbox{(scalar)} \\
    \omega & \approx & 0.36593(1+0.02179\varepsilon-0.00012\varepsilon^2)-0.19397\imo(1-0.00054\varepsilon+0.00134\varepsilon^2) & \qquad\mbox{(Dirac)} \\
    \omega & \approx & 0.49653(1+0.02251\varepsilon+0.00000\varepsilon^2)-0.18498\imo(1+0.00244\varepsilon+0.00101\varepsilon^2) & \qquad\mbox{(Maxwell)} \\
  \end{array}
\end{equation}
\end{widetext}
Thus, the real oscillation frequency indeed grows almost proportionally to $\varepsilon$, as can be concluded from smallness of the coefficients in front of $\varepsilon^2$.
The damping rates, given by absolute values of imaginary parts of the frequencies, are not monotonic function of $\varepsilon$, having a minimum
at some negative $\varepsilon$ (different for each type of perturbation) corresponding to the least damping. Let us note, that as the potential barrier is monotonically growing when $\varepsilon$ increases (fig.~\ref{potentialsQNMs}), some growth of $\re{\omega}$ (certainly, not necessarily linear) as a function of $\varepsilon$ was expected.

\begin{figure*}
\resizebox{\linewidth}{!}{\includegraphics*{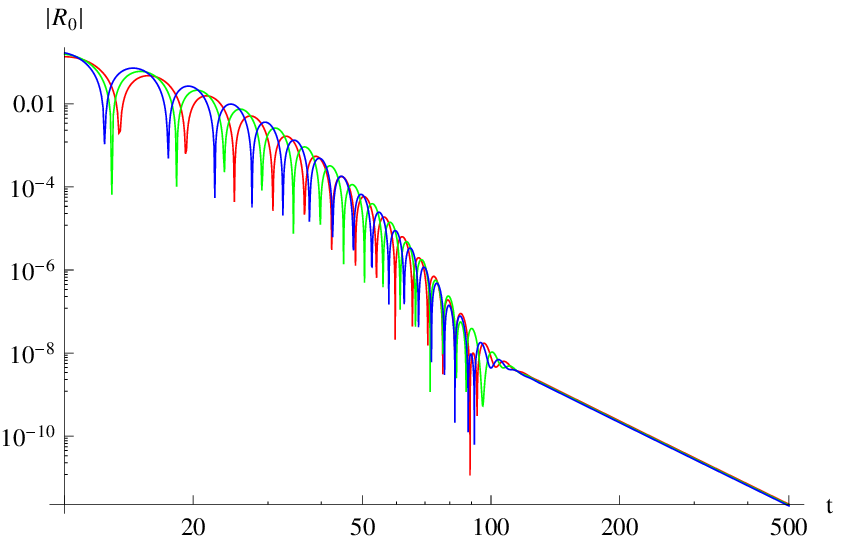}\includegraphics*{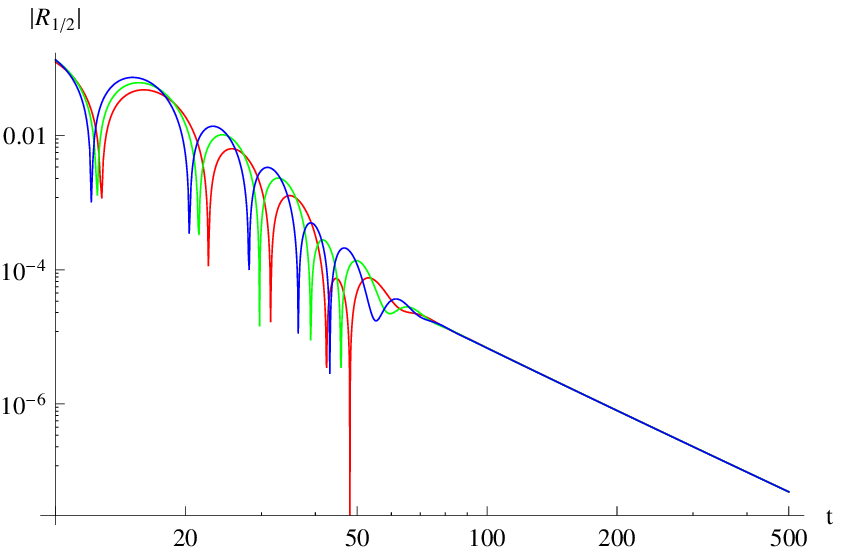}\includegraphics*{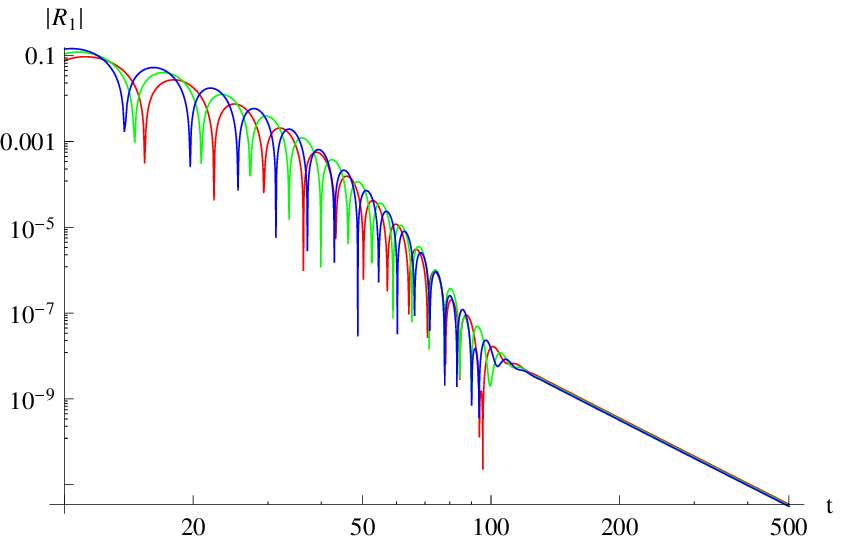}}
\caption{Time-domain profiles from left to right: for the scalar ($\ell=1$), Dirac ($\ell=1/2$), and Maxwell ($\ell=1$) fields for $\varepsilon=-4$ (red), $\varepsilon=0$ (green) and $\varepsilon=4$ (blue). Late-time tails decay law does not depend on $\varepsilon$.}
\label{tails}
\end{figure*}

The above numerical data obtained by the Frobenius method can be also checked by the time-domain integration (see Fig.~\ref{tails}).
By fitting the time-domain profile we are able to reproduce the fundamental modes obtained in the frequency domain within any desired accuracy. The asymptotic late-time tails coincide with the Schwarzschild one \cite{Price:1972}
\begin{equation}
|\Psi| \propto \left\{
                 \begin{array}{ll}
                   t^{-2\ell-3}, & \hbox{scalar, Maxwell;} \\
                   t^{-2\ell-2}, & \hbox{Dirac;}
                 \end{array}
               \right.
\quad t \rightarrow \infty.
\end{equation}
As time-domain profiles include contributions from all modes, this also proves the stability of fields in the in the Johannsen-Psaltis background.

The WKB method is known to be accurate in the regime of large multipole numbers and small overtones $n \ll \ell$.
Therefore, for $\ell\gg1$ the eikonal approximation allows us to find simple analytic expressions for
QN frequencies. Expanding the location of the maximum of the effective potential $r_{0}$ in powers of $1/\ell$ and $\varepsilon$ and making use of this expansion in WKB formula (\ref{WKBformula}) at the first WKB order and lowest orders of $\varepsilon$, we have found the following expression:
\begin{equation}
\omega=\frac{1}{3M\sqrt{3}}\left(\Re(\varepsilon)\left(\ell+\frac{1}{2}\right)-\imo\Im(\varepsilon)\left(n+\frac{1}{2}\right)\right)+{\cal O}\left(\frac{1}{\ell}\right),
\label{eikonal}
\end{equation}
where
$$\Re(\varepsilon)=1+\frac{\varepsilon}{54}+\frac{\varepsilon^2}{2916}+o(\varepsilon^2), \qquad \Im(\varepsilon)=1+\frac{\varepsilon^2}{648}+o(\varepsilon^2).$$
At large $\ell$ the data given by the above ekinal formula are in excellent concordance with the numerical one. It is well known, that the eikonal formula works well already for moderate values of $\ell = 2,3,4$, ($n=0$). Comparing the analytical expansion (\ref{eikonal}) with (\ref{eexp}) one can observe that the eikonal formula gives quite a good approximation for the the real part of $\omega$ as a function of $\varepsilon$ already for $\ell=1$ ($\ell=1/2$ for the Dirac field).

It is well known that in the eikonal approximation quasinormal frequencies of fields minimally coupled to gravity, that is, propagating in the black hole background, are determined by the centrifugal term of the effective potential, which is independent on the spin of the field.
Therefore, as a rule, the eikonal expression for $\omega$ (\ref{eikonal}) does not depend on spin as well.
Thus, we could speculate that there may be a kind \emph{universal dependence on the deformation parameter} for minimally coupled to gravity fields of any spin, when $h$ is small: the imaginary part of the fundamental quasinormal mode remains relatively close to that of the Schwarzschild black hole, while the real part grows proportionally to $\varepsilon$. This might be true also for gravitational perturbations, if the lowest dynamical mode $\ell =2$, as it takes place in General Relativity. However, if other fields are coupled to gravity, such as the dilaton, a spherically symmetric dynamical mode $\ell =0$ appears in the spectrum, which evidently cannot be described by the approximation of geometrical optics.

It is evident that gravitational perturbations cannot be analyzed within the Johannsen-Psaltis approach, as for such linearization one needs to have the stress-energy tensor at hand, whose form depends on an alternative theory under consideration.

\section{Hawking radiation}

\begin{figure*}
\resizebox{\linewidth}{!}{\includegraphics*{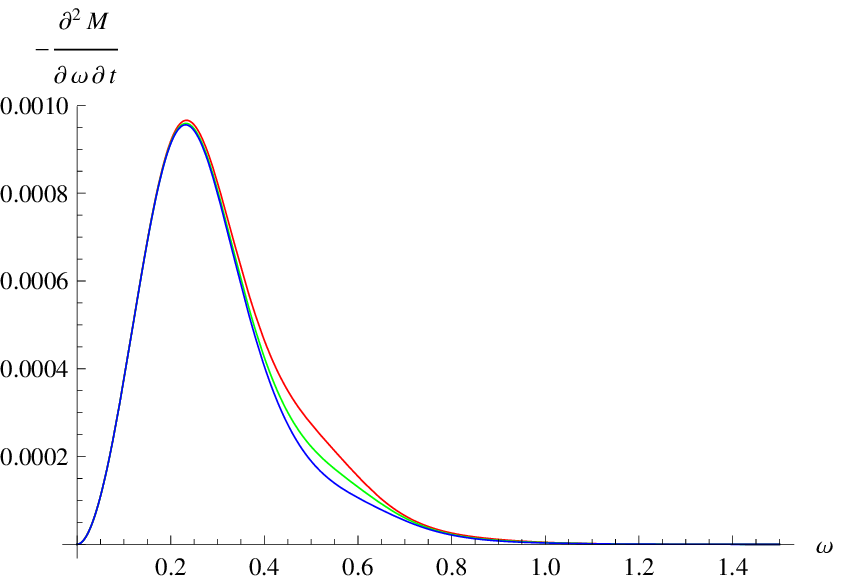}\includegraphics*{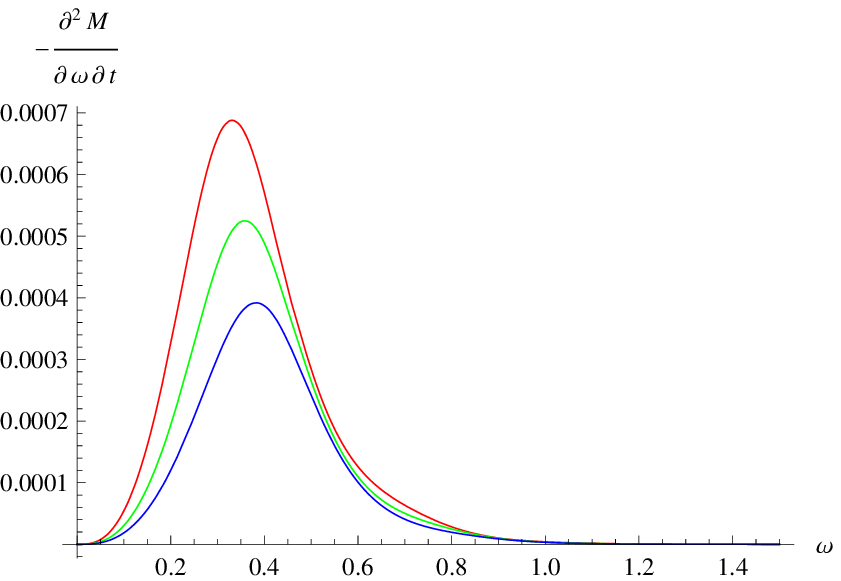}\includegraphics*{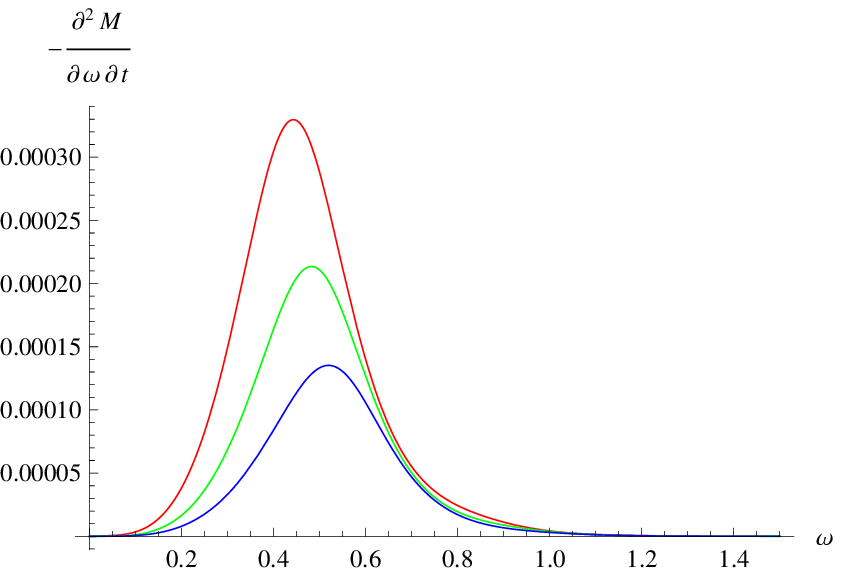}}
\caption{Energy-emission rate due to radiation from left to right: of the massless scalar, Dirac, and Maxwell (one of the spiralities) particles for $\varepsilon=-4$ (red, top), $\varepsilon=0$ (green) and $\varepsilon=4$ (blue, bottom).}
\label{HR}
\end{figure*}

Here we assume that the black hole is in the thermal equilibrium with its surroundings in the following sense: It is supposed that the black hole temperature does not change between the emission of two consequent particles, or, in other words, the system forms the canonical ensemble.

Once the coefficients $Z_{in}$ and $Z_{out}$ are calculated with the help of the shooting method related in the Sec V. B, one can find the absorption probability
\begin{equation}\label{absorption}
|{\cal A}_{\ell} |^2=1-|Z_{out}/Z_{in}|^2.
\end{equation}
for scalar and Maxwell fields. In order to calculate the absorption probability for massless Dirac particles we follow \cite{Creek:2007tw}:
\begin{equation}
|{\cal A}_{\ell} |^2=1-\frac{4\omega^2}{\overline{\lambda}}\left|\frac{Z_{out}}{Z_{in}}\right|^2.
\end{equation}

In the semi-classical approximation the asymptotically flat black hole creates and radiates particles with thermal spectrum \cite{Hawking:1974sw}.
The energy-emission rate is proportional to the absorption probability $|{\cal A}_{\ell} |^2$ \cite{Hawking:1974sw} (see also \cite{Page:1976df}):
\begin{equation}\label{energy-emission}
-\frac{d M}{dt}=\int\frac{d\omega}{2\pi}\sum_{\ell=s}^\infty\frac{(2\ell+1)|A_\ell|^2\omega}{\exp(\omega/T_H)\pm1}
\end{equation}
for particles of each spirality, where $T_H=1/8\pi M$ is the Hawking temperature. The infinite sum in (\ref{energy-emission}) converges quickly, so that in practice we need to sum over few lowest multipoles $\ell$ (note that for fermions we sum over half-integers $\ell$ and take ``+'' sign in the denominator).

\begin{table}
\caption{Energy-emission rates due to radiation of massless fields.}\label{EER}
\begin{tabular}{|r|r|r|r|}
\hline
$\varepsilon$&scalar&Dirac&Maxwell\\
\hline
$-4$&$0.000312$&$0.000213$&$0.000102$\\
$-2$&$0.000304$&$0.000187$&$0.000083$\\
$ 0$&$0.000298$&$0.000163$&$0.000067$\\
$ 2$&$0.000292$&$0.000143$&$0.000055$\\
$ 4$&$0.000288$&$0.000125$&$0.000044$\\
\hline
\end{tabular}
\end{table}

The results of computations for the energy-emission rates are shown on Fig.~\ref{HR}. The larger the absolute value of  $\varepsilon$ is, the more suppressed emission rates are. On table~\ref{EER} one can see that the deviation from Schwarzschild values are larger for fields of higher spin. This can be easily understood from the expression for the effective potentials of scalar, Maxwell, and Dirac fields (\ref{effpot}). Since $\overline{\lambda}$ always appears with the factor $k$, which contains $\varepsilon$, the larger multipole number $\ell$ is, the bigger deviation arises in the potential due to $\varepsilon$. At the same time the effective potential for the scalar field does not depend at all on $\varepsilon$ when $\ell=0$, which corresponds to the dominant contribution in the emission rates.

\section{Conclusions}

Investigation of potentially observable phenomena around black holes in a large number of alternative theories of gravity would be a difficult, if not never ending, task, since even solutions for black hole metric are not known in many of such theories. A kind of phenomenologically unified description of black holes by the Johannsen-Psaltis space-time gives us the generic model which can be studied with simple numerical tools.
Here we have analyzed a number of radiation processes in the vicinity of the Johannsen-Psaltis space-time through consideration of their essential characteristics:
\begin{enumerate}
  \item the binding energy of a particle spiralling into the black on an equatorial orbit,
  \item proper (quasinormal) modes, late-time tails and stability of scalar, Dirac, and Maxwell fields,
  \item intensity of the Hawking radiation of these fields.
\end{enumerate}

We have shown that the binding energy is increasing for more prolate than Kerr configurations, what corresponds to positive deformation parameter $\varepsilon$. In a similar manner, real oscillation frequencies of quasinormal modes are larger for positive $\varepsilon$, though, at asymptotically late time the power-law tails coincide with the Schwarzschild ones. We were able to find analytic expressions for the frequency  in two regimes: 1) for the fundamental modes and small and moderate $\varepsilon$ and 2) in the eikonal regime $\ell\gg1$. We have found that the real oscillation frequencies are approximately linear in $\varepsilon$, for small and moderate values of $\varepsilon$. Stability of the fields under consideration has been proved. The intensity of the Hawking radiation is shown to be suppressed for positive $\varepsilon$ and enhanced for negative ones.

Our work could be complemented by the analysis of classical and quantum radiation processes for massive fields. A drawback of our approach is in the apparent impossibility to decouple variables in the Boyer-Lindquist coordinates in the field equations for a non-zero rotation parameter, so that quasinormal modes and the Hawking radiation were analyzed only for non-rotating black holes. Perhaps, a kind of prolate coordinate system, which takes into account the symmetry of the space-time, could remedy the situation. Thus, a further study of properties of the Killing-Yano tensor for the Johannsen-Psaltis space-time is appealing. At the same time, strong dependence of the QNMs and intensity of Hawking radiation on the deformation parameter $\varepsilon$ leaves us hope that the obtained here results will remain qualitatively the same at least for slow rotations.

\section*{Acknowledgments}
R.~K. was supported by the European Commission grant through the Marie Curie International Incoming Program.
A.~Z. was supported by Conselho Nacional de Desenvolvimento Cient\'ifico e Tecnol\'ogico (CNPq).

\end{document}